\documentstyle[aps,multicol,epsf]{revtex}

\begin{document}
\draft

\title{A parity conserving dimer model with infinitely many absorbing states}

\author{M.C. Marques and J.F.F. Mendes}

\address{Centro de F\'\i sica do Porto and Departamento de F\'\i sica,
Universidade do Porto, \\
Rua do Campo Alegre 687, 4150 Porto, Portugal}

\maketitle

\begin{abstract}
We propose and study a model where, for the first time, two aspects are present: 
parity conservation and infinitely many absorbing states. 
Whereas steady-state simulations show that the static critical behaviour is not 
affected by the presence of multiple absorbing configurations, the influence of 
the initial state associated with the presence of slowly decaying memory effects 
is clearly displayed in time dependent simulations. 
We report results of a detailed investigation of the dependence of critical spreading 
exponents on the initial particle density.
\end{abstract}

\begin{multicols}{2}

\narrowtext

Various nonequilibrium models exhibiting phase transitions from an active to 
an absorbing phase have been extensively investigated over recent years
\cite{ann,liggett,grass1,dickman1}. 

Most of them \cite{ziff,dickman2,jensen1,bidaux,ben,takayasu} 
have been shown to belong to the universality class 
of directed percolation (DP) \cite{kinzel}. 
These studies gave support to the conjecture made 
by Janssen and Grassberger \cite{janssen,grass2} 
that continuous transitions to a unique absorbing state generically fall in
the DP class. The same applies, as far as the static critical behaviour is 
concerned, to various models with multiple absorbing states
\cite{jensen2,jensen3,mendes}. However, the dynamic critical properties
of these systems are found to be non-universal: in fact, spreading critical
exponents depend on the initial density and obey a generalized
hyperscaling relation derived by Mendes et al. \cite{mendes}. Some systems,
first claimed to have a different critical behaviour, were later included in
the DP class, upon a more careful analysis \cite{bidaux,ben}.

For sometime the only accepted exceptions to the DP universality class were
the models $A$ and $B$ introduced by 
Grassberger and collaborators \cite{grass3,grass4}:
they both have a doubly degenerated absorbing state and modulo 2 
parity conservation of kinks (-$00$- and -$11$-s). Since then, a 
growing number of models whose critical behaviour falls in 
the parity conserving (PC) universality
class have been studied. One of the first to be considered was the branching
annihilating random walk with an even number $m$ of 
offsprings (BAW) \cite{takayasu,jensen4,zhong,tauber}: in this case, 
the parity of the particles is conserved 
and there is a single absorbing state. Another 
system extensively studied is the nonequilibrium Ising model with combined 
spin-flip and spin-exchange dynamics 
(NEKIM) \cite{nora}; here, the absorbing state 
is symmetrically doubly degenerated
and the kinks have local parity conservation, similarly to what happens with 
the $A$ and $B$ models referred above.  Other models have since been discovered
and shown to display PC critical behaviour: the interacting monomer-dimer
model \cite{kim} and generalizations of the Domany-Kinzel automaton with
$n$ equivalent absorbing states \cite{haye}.  In these models, parity is not
 strictly conserved but there is complete symmetry among absorbing states;
actually, if this symmetry is broken, DP critical behaviour is recovered.

Therefore, the question about the essential feature behind PC critical
behaviour is not yet settled. Whereas in BAW models mass conservation
of modulo $2$ is determinant, in the other cases symmetry among absorbing 
states plays the relevant role. In the NEKIM model, parity conservation in
the number of kinks is not sufficient to ensure PC-like behaviour, if such
symmetry is broken\cite{nora}.

Recently, Inui and Tretyakov \cite{inui} presented a new version of a contact process 
with parity conservation, which displays a phase transition for $m=2$, 
contrarely to what happens with the original version of $BAW$. They claim, 
on the basis of Pad\'e aproximants and also numerical simulations, that 
the order parameter exponent $\beta$ should be $1$, while the values 
estimated by different authors are consistently smaller \cite{jensen4,kim,haye}.

In this Letter, we consider a model that resembles the pair-contact process
(PCP) introduced by Jensen \cite{jensen2}, but where the number of
dimers is conserved modulo $2$. Dimers cannot be generated spontaneously, 
therefore any configuration with 
isolated particles and empty sites is absorbing, and 
there is an infinite number of such configurations. This enables one to study 
the way initial configurations affect the spreading critical exponents in a 
system that belongs to the PC-universality class.

Initial conditions were shown \cite{geza1} to affect time-dependent 
critical exponents that describe the damage spreading transition (DS) 
in cases where this one coincides with the ordinary critical point; in DS, one
follows the evolution of two replicas, so this can be seen as a 
multicomponent  system, and any state where both replicas are identical is considered as absorbing.

We start by briefly reviewing the dynamic rules for the PCP.
This is a lattice model which can be described by a $2$-state
variabe $\sigma_{i}=0,1$. Nearest-neighbour pairs of particles 
(dimers) annihilate each other with probability $p$ or create,with probability
$1-p$, a particle at one of the adjacent (vacant) sites to the dimer. 
The annihilation of a pair may imply the loss of one or two other
pairs, if one (or two) of the nearest-neighbour sites to the chosen
dimer happen to be occupied; analogously when a particle is created adjacent to
a dimer, the number of pairs increases by one or two, depending on the
occupancy of the other nearest neighbour to the site that is being occupied.
It is therefore clear that there is no parity conservation in the number of 
dimers and the model displays a DP-like phase transition from an active state
with a nonzero concentration of dimers 
(for $p<p_{c}$) to a phase with infinitely many absorbing states (for $p>p_{c}$). 
Critical spreading exponents are non-universal depending upon the nature of the initial configuration.

In the one-dimensional model we now introduce, a phase with infinitely 
many absorbing states is also present. On the other hand, dynamic rules are such that there is
parity conservation in the number of dimers. This seems to be the most relevant
feature; indeed, we find that the static critical behaviour of 
this model is PC-like, thus it is not affected by the presence of multiple
absorbing states.

The basic processes are again annihilation and creation, 
which are attempted with probabilities $p$ and $1-p$, respectively. 
If one chooses to annihilate, then one dimer (represented by $\bullet=\bullet$) is 
selected at random and one looks for an adjacent dimer. If the pair of dimers
is surrounded by empty sites ($\circ$), then the two dimers are annihilated and the 
respective sites become empty 
($-\circ-\bullet=\bullet=\bullet-\circ- \rightarrow -\circ-\circ-\circ-\circ-\circ-$); 
otherwise, annihilation is produced by simply 
vacating the site that is shared by the two dimers, leaving the other sites unchanged ($-\circ-\bullet=\bullet=\bullet=\bullet- \rightarrow -\circ-\bullet-\circ-\bullet=\bullet-$). 
In this way, the number of dimers is conserved modulo $2$.
In case of a creation attempt, then if the nearest and next-nearest-neighbour 
sites of a selected dimer are, respectively, vacant and occupied, a pair of 
dimers is produced by simply occupying that vacant site 
($-\bullet=\bullet-\circ-\bullet-\circ- \rightarrow -\bullet=\bullet=\bullet=\bullet-\circ-$); 
another possibility requires the presence of three vacant sites adjacent to a dimer, 
in which case a pair of dimers is created by filling the nearest and next-nearest 
neighbour sites and leaving the third vacant site unchanged ($-\bullet=\bullet-\circ-\circ-\circ- \rightarrow -\bullet=\bullet=\bullet=\bullet-\circ-$).
In order that the presence of isolated particles may result in the enhancement of 
possible growth, we introduce a parameter $\alpha$, such that the rate of creation 
by the latter process is a fraction $\alpha$ of the total creation rate. 
In the present work, we have fixed $\alpha = 1/2$.

With these rules, diffusion is only indirectly present. One can of course enlarge
the parameter space to include diffusion, but we have not done it in the 
present work. In the absence of such processes, 
creation of dimers is inhibited in 
configurations where a pair of vacant sites is
surrounded by occupied sites. Initial configurations have therefore been 
chosen such that this situation does not occur
in the dynamical process. Also, the annihilation of isolated dimers is here possible only
through sucessive processes of creation and annihilation. This affects 
particularly (and slows down) the dynamical processes leading to absorbing 
configurations.

The order parameter of the system is the concentration of dimers, which
vanishes algebraically as $p$ approaches the critical probability $p_{c}$:
\begin{equation}
\rho \sim (p_{c}-p)^{\beta}
\end{equation}
where $\beta$ is the order parameter exponent.

Simulations were started with a fully occupied lattice and the concentration 
averaged over a long period of time once the steady state has been reached.

In fig.$1$ we show a log-log plot of the steady state concentration as a
function of $p_{c}-p$ for system size $L=2000$; time varied from $t=5000$ to
$t=2 \times 10^{5}$ MCS closest to $p_{c}(=0.2990(10))$ and 
we averaged over around $1000$
independent samples which had not yet entered the 
absorbing state. From the slope of the 
data we estimate $\beta=0.98(5)$, a value which agrees with the results of \cite{inui} and is slightly 
above the values obtained by other authors for models in the PC universality class.
Determining the critical point by steady state simulations requires an accurate
measurement of the order parameter near criticality, which becomes rather difficult
due to the critical slowing down; a scaling region of over one decade in 
$\Delta = p_{c}-p$ can clearly be observed in 
our data and leads to the above estimates for $p_{c}$ and $\beta$.

We have complemented this study with a finite-size scaling analysis based on
the ansatz that the order parameter depends on system size $L$ through the 
ratio of $L$ and the correlation length $\xi \sim \Delta^{-\nu_{\perp}}$:
\begin{equation}
\rho(p,L) \sim L^{-\beta/\nu_{\perp}} f(\Delta 
L^{1/\nu_{\perp}}) 
\end{equation}
(with $f(x) \propto x^{\beta} $ for $x\rightarrow \infty$, so that (1) 
is recovered when $L\rightarrow
\infty$).

Accordingly,
\begin{equation}
\rho(p_{c},L) \propto L^{-\beta/\nu_{\perp}}
\end{equation}

In Fig.~2, we show a log-log plot of $\rho(p_{c},L)$ as a function of L; a straight line is
obtained, from the slope of which we estimate 
$\beta / \nu_{\perp} =0.54(4)$, to  be compared with $0.48$ and $0.50$ as 
obtained in \cite{kim} and \cite{jensen4}, respectively.

We also report results of critical spreading, i.e. the evolution of the critical
system from a configuration that is very close to an absorbing state.
In the present model, there are many absorbing configurations and some
critical spreading exponents are indeed dependent upon the density of particles
in the initial state.
The quantities that are usually considered in these studies are the surviving
probability $P(t)$, the number of dimers $n(t)$ averaged over all runs, and
the mean-square distance of spreading $R^{2}(t)$ averaged over the surviving
runs. At criticality, these quantities obey, in the long time limit, 
$P(t) \sim t^{-\delta}$,  $n(t) \sim t^{\eta}$ and $R^{2}(t) \sim t^{z}$, 
and the corresponding exponents can be obtained from the straight lines shown
in double-logarithmic plots of the quantities against time. 
More precise estimates are usually obtained by looking at local slopes, for example
$-\delta (t)=ln[P(t)/P(t/m)]/ln(m)$. In a plot of the local slopes vs $1/t$, 
the critical exponents are given by the intercept of the curves for $p_{c}$ with the 
vertical axis, whereas curves for 
$p>p_{c}$ ($p<p_{c}$) veer downward (upward). This often constitutes a rather
accurate method for the determination of $p_{c}$.
It is certainly the case when one or just a few absorbing states are present
\cite{kim,dickman3}. Also in some systems with infinitely many
absorbing states, the critical point was shown not to depend on the initial
configuration and can therefore be located by the procedure we just 
described \cite{mendes}. In other cases, indications of a slight dependence 
of the critical point (as determined by the above time-dependent analysis) 
on the initial configuration have been found \cite{dickman4,geza2} and can be 
atributed to slowly decaying memory effects \cite{grass5}. 
This seems to occur also in the present case.

In these simulations we started the system in a way that a sublattice of 
alternating sites is vacant except for one central site, and the sites of the 
other sublattice are occupied with probability $q$; 
the two nearest neighbours of the central site are also occupied, 
thus constituting a perturbation of exactly two dimers.
The size of the lattice was chosen such that the spreading never hits the
boundaries. For different $q$ values, a number of independent runs, typically $10^{7}$,
were performed, up to $4000$ time steps each, and various values of $p$. 

A local slope analysis for $q=0.43$ leads to the estimate $p_{c}=0.2995(5)$,
which is fully consistent with the $p_{c}$ estimate from steady-state simulations, above;
the estimates for the dynamic scaling exponents are $\eta'=0.005(5)$,
$\delta'=0.291(6)$, $z'/2 = 0.570(6)$ and agree well with previous results 
for models in the PC universality class \cite{jensen4} ($\eta=0.000(1), \delta=
0.285(2), z/2 = 0.571(1)$). This is similar to what happens in the pair-contact process: DP
time-dependent exponents are found if one studies spreading of a perturbation to 
the "natural" configuration, the one spontaneously
generated by the critical dynamics. The estimate $q^{nat} = 0.43(2)$ ($q^{nat}$ meaning the value of $q$ that corresponds to the density in the natural configuration) is in good agreement, 
within numerical accuracy, with the one obtained by generating the natural configuration in samples  of
size $L=2000$, starting with a full lattice and using periodic boundary conditions. 

For different $q$ values, power laws still hold, but with different critical spreading
exponents, and a slight shift of the critical point. In Fig.Ä3 we show the 
local slope analysis for $q=0.2$, leading to the following estimates: 
$\delta' = 0.366(6), \eta' = -0.069(5), z'/2 =0.567(6), p'_{c}=0.3000(5)$. 
The generalized hyperscaling relation $\eta' + \delta' - z'/2 =  - \delta$ 
\cite{mendes} is well satisfied by these values. 

Our results for the $q$-dependence of  $p'_{c}, \delta',\eta'$ and $z'/2$ are given in Table 1.
A check of the generalized hyperscaling is also included. Indeed, the exponent
$z'$ does not present a significative dependence on the parameter $q$, given the numerical errors.
Then, as expected, the exponent governing the population growth in surviving critical trials, 
$\delta'+\eta'$, does not depend on the initial particle concentration.

In conclusion, we have performed a similar study to what has been done before
\cite{jensen2} for the PCP (and other models with multiple absorbing states \cite{mendes}) 
but now a system with parity conservation (in the number
of dimers) is for the first time investigated. A field theory formalism 
to appropriately describe this situation is now required.

Whereas the presence of multiple absorbing states does not affect the static behaviour,
spreading critical behaviour is expressed by power laws whose exponents
depend on the initial particle density ($=q/2$). There is also
a monotonic shift in $p'_{c}$, which, when found in other systems, has been
atributed \cite{geza2} to slowly decaying memory effects displayed by the
non-order field. The generalized hyperscaling relation  is verified.

A possible mapping of this system to a SOC model\cite{dickman5} is currently under study.  

This work was supported partially by PRAXIS XXI (Project /2/2.1/Fis/299/94) and  
NATO (Grant CRG-970332). We thank Ron Dickman and Maria Augusta Santos for useful 
discussions and critical reading of the manuscript. We also thank G\'eza \'Odor 
for a stimulating discussion in the initial stage of this work.

\end{multicols}

% FIG 1
\begin{figure}
\epsfxsize=60mm
\epsffile{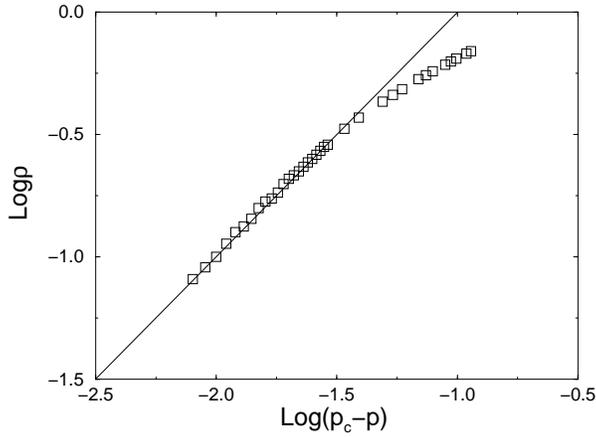}
\caption{Log$_{10}$-log$_{10}$ plot of the concentration of dimers, $\rho$ versus $p_c - p$, 
with $p_c = 0.2990$.}
\end{figure}

% FIG 2
\begin{figure}
\epsfxsize=65mm
\epsffile{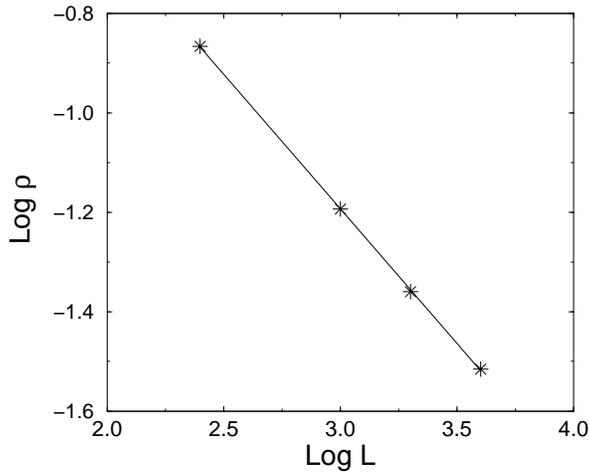}
\caption{Log$_{10}$-log$_{10}$ plot of $\rho(p_{c}, L)$ versus $L$ ($L=250$, $500$, $1000$ and $2000$).}
\end{figure}

% FIG 3
\begin{figure}
\epsfxsize=120mm
\epsffile{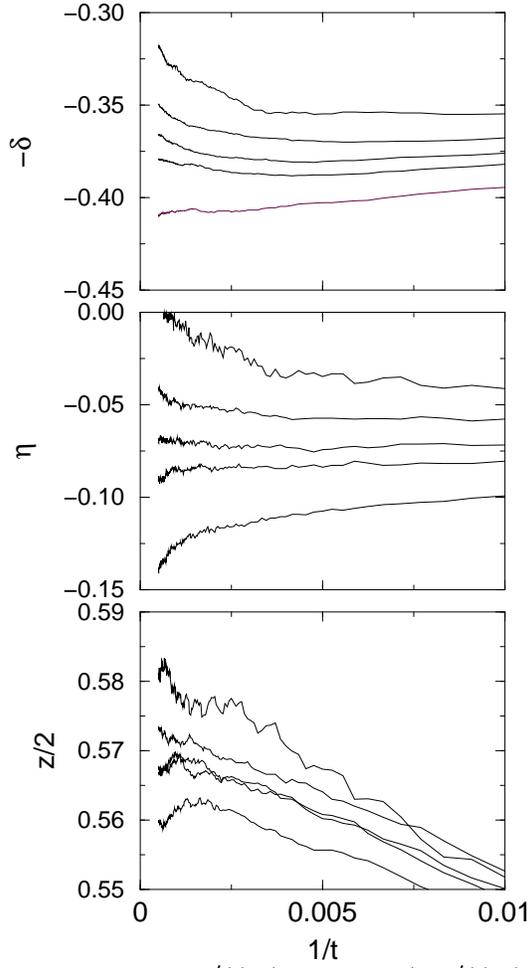}
\caption{Local slopes $-\delta'(t)$ (upper panel), $\eta'(t)$ (midle-panel) and 
$z'/2(t)$ (lower panel) for $q=0.2$. Each panel contains five curves with, from bottom to top, $p=0.2980$, $0.2990$, $0.3000$, $0.3010$ and $0.3020$.}
\end{figure}

% TABLE 1
\begin{center}
\begin{tabular}{cccccc}
\hline \hline
$q$      & $p'_{c}$        & $\delta'$   & $\eta'$    & $z'/2$    & $\eta' + \delta' - z'/2  + \delta$ \\
\hline \\
DP        & -              & 0.1597(3)   & 0.312(2)   & 0.632(1)  & 0.000(1)  \\
PC        & -              & 0.285(2)    & 0.000(1)   & 0.571(1)  & -0.001(1) \\
\hline
0.1       & 0.3005(5)      & 0.400(6)    & -0.106(5)  & 0.570(6)  & 0.01(2) \\
0.2       & 0.3000(5)      & 0.366(6)    & -0.069(5)  & 0.567(6)  & 0.02(2) \\
0.43      & 0.2995(5)      & 0.291(6)    &  0.005(5)  & 0.570(6)  & 0.01(2) \\
0.45      & 0.29935(5)     & 0.284(6)    &  0.014(5)  & 0.570(6)  & 0.01(2) \\
0.5       & 0.2990(5)      & 0.271(6)    &  0.033(5)  & 0.572(6)  & 0.02(2) \\
0.6       & 0.2990(5)      & 0.239(6)    &  0.062(5)  & 0.573(6)  & 0.01(2) \\
0.8       & 0.2985(5)      & 0.181(6)    &  0.126(5)  & 0.576(6)  & 0.02(2) \\
0.9       & 0.2985(5)      & 0.156(6)    &  0.154(5)  & 0.579(6)  & 0.02(2) \\
\hline \hline 
\end{tabular}
\end{center}
TABLE CAPTION: The table shows the $q$-dependence of the critical 
parameter $p'_c$ and critical exponents $\delta'$, $\eta'$ and $z'/2$. A test 
of generalized hyperscaling is also included. 

\end{document}